\documentclass[conference]{IEEEtran}
\IEEEoverridecommandlockouts
\usepackage{amsmath,amssymb,amsfonts}
\usepackage{mathtools}
\usepackage{algorithmic}
\usepackage{graphicx}
\usepackage[sorting=none]{biblatex}
\usepackage{hyperref}
\usepackage{booktabs}
\usepackage{textcomp}
\usepackage{xcolor}

\usepackage{etoolbox}
\makeatletter
\patchcmd{\@makecaption}
  {\scshape}
  {}
  {}
  {}
\makeatother
\usepackage{caption}
\usepackage{fancyhdr} 
  
\begin{document}

\title{Computational Empathy Counteracts the Negative Effects of Anger on Creative Problem Solving\\
\thanks{The authors would like to acknowledge funding from the MIT Media Lab member companies.}
}

\author{\IEEEauthorblockN{Matthew Groh}
\IEEEauthorblockA{\textit{MIT Media Lab} \\
Cambridge, MA \\
groh@mit.edu}
\and
\IEEEauthorblockN{Craig Ferguson}
\IEEEauthorblockA{\textit{MIT Media Lab} \\
Cambridge, MA \\
fergusoc@media.mit.edu}
\and
\IEEEauthorblockN{Robert Lewis}
\IEEEauthorblockA{\textit{MIT Media Lab} \\
Cambridge, MA \\
roblewis@media.mit.edu}
\and
\IEEEauthorblockN{Rosalind W. Picard}
\IEEEauthorblockA{\textit{MIT Media Lab} \\
Cambridge, MA \\
picard@media.mit.edu}
}

\maketitle
\thispagestyle{fancy}
\begin{abstract}

How does empathy influence creative problem solving? We introduce a computational empathy intervention based on context-specific affective mimicry and perspective taking by a virtual agent appearing in the form of a well-dressed polar bear. In an online experiment with 1,006 participants randomly assigned to an emotion elicitation intervention (with a control elicitation condition and anger elicitation condition) and a computational empathy intervention (with a control virtual agent and an empathic virtual agent), we examine how anger and empathy influence participants' performance in solving a word game based on Wordle. We find participants who are assigned to the anger elicitation condition perform significantly worse on multiple performance metrics than participants assigned to the control condition. However, we find the empathic virtual agent counteracts the drop in performance induced by the anger condition such that participants assigned to both the empathic virtual agent and the anger condition perform no differently than participants in the control elicitation condition and significantly better than participants assigned to the control virtual agent and the anger elicitation condition. While empathy reduces the negative effects of anger, we do not find evidence that the empathic virtual agent influences performance of participants who are assigned to the control elicitation condition. By introducing a framework for computational empathy interventions and conducting a two-by-two factorial design randomized experiment, we provide rigorous, empirical evidence that computational empathy can counteract the negative effects of anger on creative problem solving.

\end{abstract}


\section*{Introduction}

Empathic virtual agents are artificial intelligence systems designed to perceive and express affect in order to simulate the appearance of empathy in interactions with humans. Computational empathy involves recognizing an individual's emotional state and responding appropriately via affective mimicry and perspective taking~\cite{paiva2017empathy}. While affective computing~\cite{picard2000affective} seeks to address the challenges of recognizing emotions and responding empathically, these are not solved problems and there remain many open questions on how to evaluate computational empathy~\cite{yalcin_evaluating_2019}. In evaluating empathy in humans, psychologically validated methods like the Interpersonal Reactivity Index~\cite{davis1983measuring}, Empathy Quotient~\cite{baron2004empathy, lawrence2004measuring}, and Toronto Empathy Questionnaire~\cite{spreng2009toronto} involve measuring the self-reported empathy traits and preferences of an individual, but these first-person scales are not relevant for evaluating how individuals perceive empathy expressed by others. In order to evaluate perceived empathy, recent evaluations have transformed previously validated methods into evaluations by outside observers, which can be either an interaction partner or a third party~\cite{yalccin2019evaluating, hastie2016remember}. However, these evaluations by outside observers can be affected by a range of factors including observer-level factors (sociocultural background and experience with computers), context-level factors (the role of the agent as a companion or trainer and the quality of experience from a perspective of effectiveness, efficiency, utility, and acceptability), and agent-level factors (likeability, anthropomorphism, animacy, perceived intelligence and safety)~\cite{yalcin_evaluating_2019}. 

Rather than evaluating how empathic a virtual agent appears, we focus this paper on two contributions: (1) the introduction of a constrained context and a virtual agent designed to respond to humans with contextually appropriate affective mimicry and perspective taking and (2) the evaluation of the effectiveness of such an empathic virtual agent in enhancing a human's cognitive performance~\cite{maes1995agents}. Motivated by the question, ``How does computational empathy influence creative problem solving?,'' we evaluate how an empathic virtual agent, which is integrated into an online word guessing game, Affective Wordle Lab (based closely on Wordle~\cite{wordlenyt}), influences cognitive performance.

Many research experiments have shown that emotions influence creative problem solving and decision-making. For example, past research has experimentally demonstrated that positive affect (as elicited in the late 1980s by either a short blooper reel or a small gift of candy) facilitates creative problem solving~\cite{isen1987positive}. This research operationalized creative problem solving based on two tasks, Duncker's Candle task~\cite{duncker1945problem} and the Remote Associates Test (RAT) ~\cite{mednick1964incubation}, which involve finding the ``relatedness in diverse stimuli that normally seem unrelated''~\cite{isen1987positive} and require ``breaking set'' -- e.g., recognizing the box of tacks in Duncker's Candle task as a box and tacks and recognizing relatedness of words in the RAT based on many different kinds of associations. In contrast to the effects of positive affect on creative problem solving and decision-making~\cite{isen1993positive}, past research on experimentally elicited anger shows anger inhibits decision-making by reducing depth of processing and increasing reliance on heuristic processing ~\cite{lerner_beyond_2000,lerner_heart_2004, lerner_portrait_2006, lerner_emotion_2015}. 

\section*{Related Work}

\subsection*{Video Games, Emotions, and Cognitive Testing}

Video games are an area of growing interest in affective computing and other computational sciences, where much of the work benefits from two key properties of video games. First, video games can deeply engage players and evoke poignant emotional experiences, thus enabling the study of various psychological constructs on large study or real-world user populations \cite{ryan2006motivational,ali2022escape,Ferguson&Lewis2021}. Second, video games can be customised to create highly controlled environments that probe specific cognitive and affective processes, thus giving researchers fine-grained control of their experiments and the ability to objectively measure constructs of interest by analyzing the game telemetry data of player actions and decisions \cite{yannakakis2014emotion, mcilroy2021detecting}. For example, numerous mini-games have been developed to assess cognitive processes such as working memory, motivation, appraisal of and aversion to risk or reward, creativity, and other general executive functions. These tests aid our understanding of the mechanisms of human decision-making and problem solving, and have found utility in various contexts including mental health monitoring where impairments in decision-making processes may be indicative of disorders like anxiety or anhedonia \cite{lumsden2016,gillan2021smartphones}. 

\subsection*{Empathic Virtual Agents}

In the past twenty years, many experiments have empirically demonstrated the power of empathic virtual agents to influence human affect including undoing negative feelings of frustration~\cite{klein2002computer}, increasing people's feelings of being cared for~\cite{bickmore2004towards, brave2005computers}, altering people's feelings of fear into neutral feelings~\cite{moridis2012affective}, and reducing public speaking anxiety~\cite{murali2021friendly}. While the fundamental tenets of affective mimicry and perspective taking drive computational empathy, the design space for computational empathy is combinatorically large. One recent study examined the systematic manipulation of animation quality, speech quality, and rendering style and their impacts on people's perceptions of virtual agents in terms of naturalness, engagement, trust, credibility, and persuasion in a health counseling domain~\cite{parmar2022designing}. In virtual agent chatbots that are limited to text interfaces, the range of possible conversations remains very large yet examples of personalized machine-learning based chatbots have been shown to interact empathically with humans and be perceived as likable~\cite{ghandeharioun_towards_2019,ghandeharioun_emma_2019,firdaus2022enjoy}. While past experiments have examined human perception of computational empathy, the authors are not aware of past experiments examining the impact of computational empathy on cognitive performance metrics.

\section*{Methods}

\subsection*{Participants}

We recruited 1,006 participants from Prolific, an online platform for recruiting research participants. We restricted recruitment to individuals on Prolific who live in the United States and speak English as a first language. As a robustness check to the inclusion criteria, we ask participants ``Are you a native English speaker?'', and 99.7\% of participants respond ``Yes.'' Participants' ages range from 18 to 84 with a median of 35, and 53\% of participants identify as female. Before participants played Wordle in this experiment, we asked ``How many times have you played Wordle'' to which 31\% of participants respond they have never played Wordle, 4\% just once, 19\% 2 to 10 times, 42\% 11 to 100 times, and 3\% have played Wordle over 100 times. 

\subsection*{Experimental Design}

We pre-registered the experiment on \textit{AsPredicted} at \href{https://aspredicted.org/yx4k3.pdf}{https://aspredicted.org/yx4k3.pdf}. 

Participants are randomly assigned to two interventions: an emotion elicitation intervention with two conditions (control and anger) and a virtual agent personality intervention with two conditions (control personality and empathic personality). In this 2x2 factorial design, we assign participants to control and treatment conditions with equal likelihoods; 26\% of participants were assigned to the control-control group, 26\% of participants were assigned to the anger-control group, 24\% of participants were assigned to the control-empathy group, and 24\% of participants were assigned to the anger-empathy group. 

\subsubsection*{Emotion Elicitation}

We based the emotion elicitation intervention on a reflective writing exercise from Small and Lerner (2008)~\cite{small_emotional_2008}. In both conditions, we ask participants to respond to two similarly structured questions with a minimum response of 150 characters each. The goal of these questions is to generate equivalent cognitive loads while activating incidental anger in one condition and not activating any specific emotion in the other condition. An incidental emotion refers to an emotion unrelated to the main task, which contrasts with an integral emotion, which refers to an emotion intrinsically tied to the main task. In this experiment, we focus on incidental anger. 

For participants assigned to the control elicitation condition, we first ask: ``What are three to five activities that you did today? Please write two-three sentences about each activity that you decide to share. (Examples of things you might write about include: walking, eating lunch, brushing your teeth, etc.)'' After they answer, we follow up with a second question: ``Now, we’d like you to describe in more detail the way you typically spend your evenings. Begin by writing down a description of your activities and then figure out how much time you devote to each activity. Examples of things you might describe include eating dinner, studying for an exam, working, talking to friends, watching TV, etc. If you can, please write your description so that someone reading this might be able to reconstruct the way in which you, specifically, spend your evenings.'' 

For participants assigned to the anger elicitation condition, we first ask: ``What are the three to five things that fill you with anger? Please write two-three sentences about each thing that fills you with anger. (Examples of things you might write about include: being treated unfairly by someone, being insulted or offended, etc.)'' After they answer, we follow up with a second question: ``Now, we’d like you to describe in more detail the one situation that makes you (or has made you) experience the most anger. This could be something you are presently experiencing or something from the past. Begin by writing down what you remember of the anger-inducing event(s) and continue by writing as detailed a description of the event(s) as is possible. If you can, please write your description so that someone reading this might even feel anger just from learning about the situation. What is it like to be in this situation? Why does it make you so feel such anger?''

\subsubsection*{Affective Wordle Lab}

After responding to the reflective writing exercise, we provide instructions to participants for playing Wordle and invite participants to play four rounds of Wordle. The rules of the Affective Wordle Lab experiment are the same as the rules in the \href{https://www.nytimes.com/games/wordle/index.html}{official version of Wordle hosted by the New York Times}~\cite{wordlenyt}. The goal of each round is to guess a 5-letter-word within 6 guesses. After each guess, players are informed whether each letter in their guess is: (a) in the correct position for the solution word, (b) in the solution word but not the correct position, or (c) not in the solution word. Players can use this information to home in on the 5-letter-word solution. Wordle closely resembles the word game Jotto, which attracted the interest of computer programmers in the early 1970s for studying information theory aspects of the game~\cite{beeler1971information}.

We adapted Wordle's game mechanics such that participants play 4 rounds and have the option to play additional bonus rounds, which stands in contrast to the official version of Wordle's standard limit of a single round per day. The list of 12,972 acceptable guesses in the Affective Wordle Lab is identical to the list of acceptable guesses in the official version of Wordle and all solutions are chosen from the official list of 2,315 acceptable solutions (the solutions are a subset of the acceptable guesses, which are based on common word use). 12,972 acceptable guesses corresponds to a possibility space of about $10^{24}$ (precisely bounded above by $12,972*12,971*12,970*12,969*12,968*12,967$) combinations of guesses for arriving at the correct 5-letter-word.

We selected four neutral words as solutions to the four rounds in the following order: ``plant,'' ``fuzzy,'' ``diner,'' and ``image.'' We expected ``plant'' and ``diner'' to be relatively easy, ``image'' to be moderately difficult because it begins with a vowel, and ``fuzzy'' to be difficult because it contains uncommon double letters. 

We hosted the Affective Wordle Lab version of Wordle at
\href{https://wordlelab.media.mit.edu/}{https://wordlelab.media.mit.edu/} based on an \href{https://github.com/cwackerfuss/react-wordle}{ open-source clone of Wordle} that we extended with a Django backend and a MySQL database. 


\begin{figure}[htbp]
\includegraphics[width=0.49\textwidth]{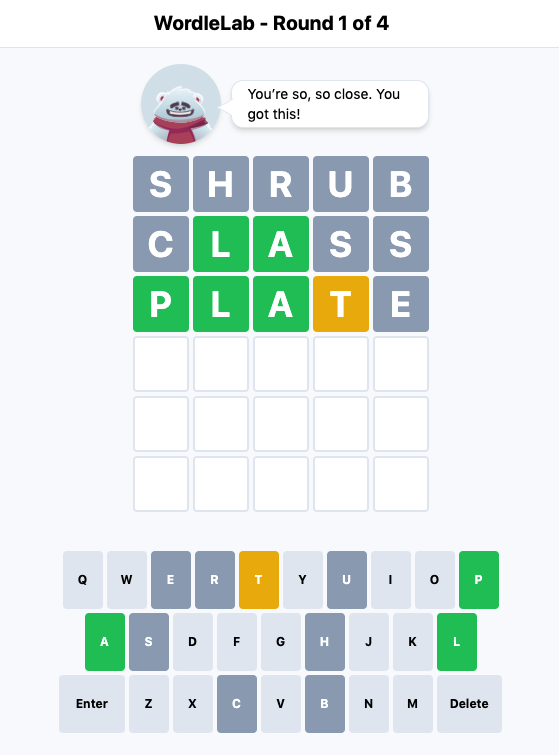}
\caption{Screenshot of the Affective Wordle Lab experiment with an empathic virtual agent. }
\label{affective-mimicry}
\end{figure}

\subsubsection*{Virtual Agent Personalities}

In a deviation from the popular Wordle game, the Affective Wordle Lab includes a virtual agent, a dynamic cartoon polar bear wearing a red scarf based on the \href{https://rive.app/community/2244-4463-animated-login-screen/}{Animated Login Screen by JcToon on Rive}. We designed the virtual agent to either display a ``control'' or ``empathic'' personality.

In the control condition, the virtual agent is always idle and is programmed to only communicate obvious game status information. Specifically, the virtual agent's speech bubble is limited to ``Guess [1-6] of 6'' for each guess iteration or ``You [won/lost] after [1-6] guesses'' between each round.

\begin{figure}[htbp]
\includegraphics[width=0.49\textwidth]{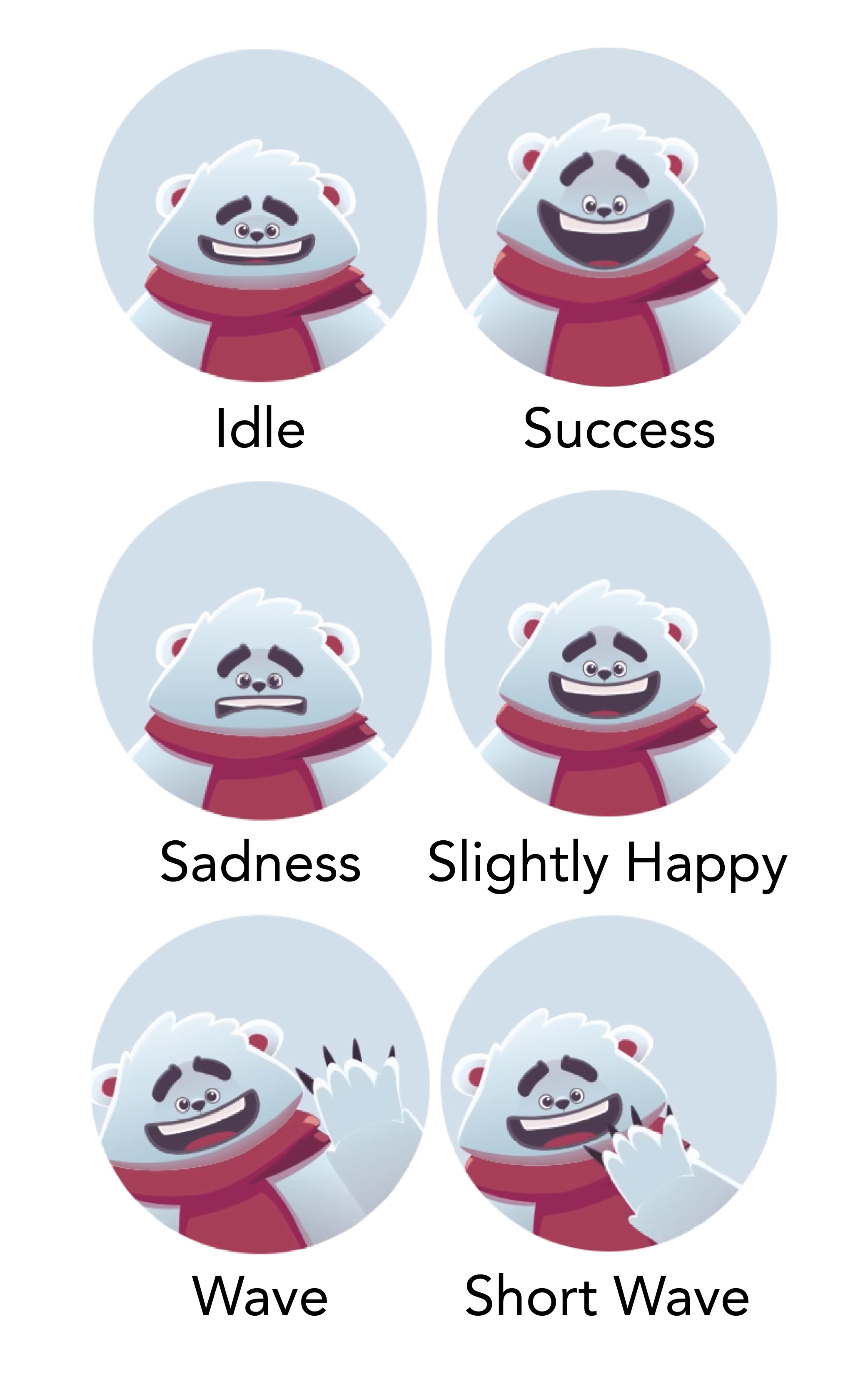}
\caption{Screenshots of the virtual agent in the six dynamic displays designed for affective mimicry. The idle display shows the bear moving his eye brows up and down every few seconds; success shows the bear shrug and burst into a wide smile with raised eyebrows; sadness shows the bear shrug, raise head, and shrug again while frowning and wiggling its ears; slightly happy shows the bear shrug and raise its head into a smile with raised eyebrows; the wave shows the bear waving buoyantly three times with raised eyebrows; the short wave shows the bear waving two times with raised eyebrows. }
\label{affective-mimicry2}
\end{figure}

In the empathy condition, the virtual agent makes expressions based on game-specific contexts to empathize with the participant via affective mimicry and perspective taking. In particular, we programmed the virtual agent to take into account how many possible words remain, how many guesses the participant has made, how quickly a participant responds,  how many letters a participant has uncovered, whether a guess is valid, whether a participant wins or loses, and whether a participant is idling. In Figure~\ref{affective-mimicry2}, we present screenshots of the virtual agent's six expressions, and in Table~\ref{context-expression-message}, we detail the 6 expressions and 39 messages paired with 13 game-specific contexts. We draw on the management science of nonverbal behavior~\cite{carney2021ten} to pair these expressions, messages, and contexts.

Each context is associated with 1 or 2 expressions paired with 1, 4, or 6 messages. The virtual agent never repeats the same message in the same round and always selects a new message for each round for contexts with 4 messages. The virtual agent selects a message based on the order of contexts presented in Table~\ref{context-expression-message}; for example, ``Fewer than 6 words remaining'' trumps ``Fast Guess (under 4 seconds),'' which trumps ``5th guess'' and so forth. 

\subsubsection*{Post-Game Questionnaire}

After participants complete 4 rounds of Wordle, we ask participants two additional sets of follow-up questions. First, we ask participants ``How are you feeling right now?'' and the experiment interface provides affective sliders~\cite{betella2016affective} for participants to report their valence and arousal. Second, we ask participants to answer 3 questions from the Cognitive Reflection Test (CRT) designed to measure depth of reflective reasoning~\cite{frederick_cognitive_2005}. After participants respond to the questions from the Cognitive Reflection Test, we congratulate participants for finishing the experiment and provide a link back to the Prolific website where participants can collect their payment. After collecting payment, participants have the option to continue playing more rounds of Wordle.

\begin{table*}[!htbp] \centering

\begin{tabular}{llll}
\toprule
                       Context &     Expression &                                          Message &                                       Message \\
\midrule
           Fewer than 6 words remaining &     Wave Short &                 You're so, so close. You got this! &                                              \\
       Fast Guess (under 4 seconds) &     Wave Short &                   Wow, you're so fast! Incredible! &                                              \\
      Slow Guess (over 60 seconds) &     Wave Short &                  Taking your time really paid off! &                                              \\
                     1st guess &     Wave Short &                           Good luck! You got this! &                 Another round! You can do this! \\
                     1st guess &     Wave Short &                       You've got the hang of this! &                    I know you can get this one! \\
                     5th guess &           Idle &           Two guesses left, that's plenty of time! & Last two guesses! Trust yourself, you got this. \\
                     5th guess &           Idle &             This is a tough one, but you're close! &     This one can be hard, but I believe in you! \\
                     6th guess &           Wave &   Just breathe and think it through. You got this! &  Stay calm and use all the facts you uncovered. \\
                     6th guess &           Wave &                   You final chance. You can do it! &       Don't give up now! Stay calm and breathe. \\
        Fewer than 101 words remaining &     Wave Short &                             You're getting closer! &        Oh nice, that really narrowed the field! \\
        Fewer than 101 words remaining &     Wave Short &                     Ooh, you're getting close now! &                   That was a really good guess! \\
   Additional letters revealed &        Success &                           Wow! What a great guess! &           Ooh nice one! I didn't think of that. \\
   Additional letters revealed &        Success &           You learned more information! Nice work! &                                    Great guess! \\
No additional letters revealed & Slightly Happy &          Okay! Well now we know what doesn't work. &                 Nice! Now we know what to avoid \\
No additional letters revealed &        Sadness &                  Aww, I was sure that would be it. &                         Hmm, what could it be?! \\
                       Invalid &        Sadness & Oops! I don't know that word! Give it another try. &                                              \\
                           Win &            Win &                        This must be your lucky day &                Two guesses?! Are you a wizard?! \\
                           Win &            Win &                 Three guesses? You're a rock star! &             Great job! You won in four guesses! \\
                           Win &        Success &               You did it! You won in five guesses! &                 That was close, but you did it! \\
                          Loss &        Sadness &                You almost had it! Let's try again. &                                             \\
Idle (triggered at 90 seconds) &     Wave Short &           It's good to think it through carefully. &                               I believe in you! \\
Idle (triggered at 90 seconds) &     Wave Short &          It's okay to feel stumped. You'll get it! &                                              \\
Idle (triggered at 90 seconds) &        Sadness &                    This one is a toughy, isn't it? &                                              \\
\bottomrule
  \\[-1.8ex] 
    \\[-1.8ex] 
\end{tabular}
\caption{The 6 expressions and 39 messages associated with 13 game-specific contexts. Each context is associated with 1 or 2 expressions and 1, 4 or 6 messages. }
\label{context-expression-message}
\end{table*}

\subsection*{Dependent Variables}

We examine four measures of game performance: (1) a binary variable for winning for each round, (2) the number of guesses per round, (3) an adjusted number of guesses per round where participants who lost are assigned 7 instead of 6 guesses, and (4) entropy reduction at the guess level, which is computed as the number of bits remaining: $log_2(w)$ where w is the mean number of words remaining for all possible solutions after each guess. We consider entropy reduction based on both a reduction of the 2,315 possible 5-letter solutions and the 12,972 possible 5-letter guesses in the official Wordle game. These are two approaches to evaluating entropy reduction of guesses, but other reasonable approaches could alternatively consider the five-letter-words from Scrabble, the Oxford English dictionary, or another source. Likewise, other reasonable approaches to evaluating entropy reduction could also take into account the word frequency. We limit our analysis to the reduction of the possible 5-letter solutions and guesses in the official Wordle game and leave additional analyses for future work.

In addition to game performance, we examine self-reported valence and arousal from the post-game questionnaire, whether participants engage in additional game play immediately after finishing the experiment, the sentiment of participants' guesses based on the VADER rule-based model~\cite{hutto2014vader}, the time between guesses, the word frequency of participants' guesses, and the number of invalid attempts submitted (e.g. a guess of ``QQQQQ'' is an example of an invalid attempt).

\section*{Results}

\subsection*{Round Level Performance}

We evaluate treatment effects of the anger elicitation intervention and the empathic virtual agent personality by running the following pre-registered ordinary least squares (OLS) regression where $Y_{i,t}$ is a dependent variable (specified above) for individual, $i$, in round $t$, A is a binary variable for assignment to the anger elicitation condition, E is a binary variable for assignment to the computational empathy intervention, $\alpha$ and $\beta_{1-3}$ are the regression intercept and coefficients, respectively, and $\epsilon$ is the error term clustered at the individual level~\cite{abadie_when_2017}: 

\begin{equation}
\begin{multlined}
Y_{i,t} = \alpha + \beta_1A_{i,t} + \beta_2E_{i,t} + \beta_3A_{i,t}E_{i,t} + \epsilon_i
\end{multlined}
\end{equation}

We find the effects of both the anger condition and the interaction between the anger and empathy condition are statistically significant at the $p<0.05$ significance level. We report these results in Table~\ref{round-results}. Relative to the control group, participants assigned to the anger condition won 7 percentage points less frequently ($p=0.021$), made an additional 0.15 guesses ($p=0.017$), and made an additional 0.21 adjusted guesses ($p=0.012$). Relative to participants assigned to the anger condition but not the empathy condition, participants assigned to both the anger and empathy condition won 8 percentage points more often ($p=0.040$), made 0.21 fewer guesses ($p=0.018$), and made 0.30 fewer adjusted guesses ($p=0.016$). We do not find assignment to the empathy intervention increases performance relative to participants assigned to the control emotion elicitation condition; participants assigned to the empathy condition won 2 percentage points less frequently ($p=0.45$), made an additional 0.11 guesses ($p=0.10$), and made 0.13 additional adjusted guesses ($p=0.143$). These results remain the same with the inclusion of round fixed effects to the linear model. 

With 1,006 participants and 4 rounds of Wordle, we should have 4,024 observations in the regression analysis, but instead we have 3,975 observations. We are missing 1.2\% of observations due to interruptions in some participant's internet connections that allowed 24 participants (2.4\% of participants) to continue the experiment without all their responses logged to the experiment's server.

\begin{table}[!htbp] 
\centering
\begin{tabular}{@{\extracolsep{5pt}}lccc}
\\[-1.8ex]\hline
\hline \\[-1.8ex]
\\[-1.8ex] & \multicolumn{1}{c}{Did Win} & \multicolumn{1}{c}{Guesses} & \multicolumn{1}{c}{Guesses (Adjusted)}  \\
\hline \\[-1.8ex]
  Constant & 0.72$^{***}$ & 4.82$^{***}$ & 5.10$^{***}$ \\
  & (0.02) & (0.04) & (0.06) \\[.1cm] 
 Anger & -0.07$^{*}$ & 0.15$^{*}$ & 0.21$^{*}$ \\
  & (0.03) & (0.06) & (0.08) \\[.1cm] 
 Empathy & -0.02$^{}$ & 0.11$^{}$ & 0.13$^{}$ \\
  & (0.03) & (0.06) & (0.09) \\[.1cm] 
 Anger * Empathy & 0.08$^{*}$ & -0.21$^{*}$ & -0.30$^{*}$ \\ 
  & (0.04) & (0.09) & (0.12) \\[.1cm] 
\hline \\[-1.8ex]
 Observations & 3,975 & 3,975 & 3,975 \\
 Number of Participants & 1006 & 1006 & 1006 \\
\hline
\hline \\[-1.8ex]
\multicolumn{4}{r}{$^{*}$p$<$0.05; $^{**}$p$<$0.01; $^{***}$p$<$0.001} \\
\end{tabular}
\caption{Ordinary least squares (OLS) regressions with robust standard errors clustered at the participant level.}
\label{round-results}
\end{table}

\subsection*{Heterogeneity of Treatment Effects on Performance}

We examine heterogeneity of treatment effects on round-level performance by including experience playing Wordle at least once before, depth of reflective reasoning as proxied by the CRT measured from 0-3, and self-reported sex~\cite{tannenbaum2019sex} in the OLS regressions.

Formally, Equation 2 includes the same terms as Equation 1 but also includes $H_{i}$, which is the heterogeneous feature of interest (either a binary variable for playing Wordle at least once before, a continuous variable from 0 to 3 indicating performance on the CRT, or a binary variable for Female). $\beta_{4}$ represents the direct association of the heterogeneous feature with the dependent variable, while $\beta_{5-7}$ are the associations of its interactions (i.e., the heterogeneous effects):

\begin{equation}
\begin{split}
Y_{i,t} = & \: \alpha + \beta_1A_{i,t} + \beta_2E_{i,t} + \beta_3A_{i,t}E_{i,t} + \\ 
          & \: \beta_4H_{i} + \beta_5H_{i}A_{i,t} +  \beta_6H_{i}E_{i,t} + \\
          & \: \beta_7H_{i}A_{i,t}E_{i,t} +\epsilon_i
\end{split}
\end{equation}

 We find that participants who had played Wordle at least once before this experiment won 20\% more frequently ($p<0.001$), took 0.35 fewer guesses ($p<0.001$), and took 0.55 fewer adjusted guesses ($p<0.001$). Likewise, we find that for every CRT question participants answered correctly, they won 9\% more frequently ($p<0.001$), took 0.20 fewer guesses ($p<0.001$), and took 0.29 fewer adjusted guesses ($p<0.001$). We do not find significant difference between men and women's performance. Moreover, we do not find statistically significant effects of interactions (i.e., $\beta_{5-7}$) between either experience playing Wordle, CRT performance, or sex and the experimental conditions on round-level performance. 
 
\subsection*{Guess Level Entropy Reduction}

We evaluate treatment effects on entropy reduction at the guess level as an additional performance metric. Specifically, we run OLS regressions following Equation 1 with an additional index $g$ on each term to denote the guess index. We measure entropy reduction by computing the $log_2(w)$ where w is the mean number of possible words (out of either the 2,315 solutions or 12,972 valid words) remaining after each guess. 

\begin{figure}[h]
\includegraphics[width=0.5\textwidth]{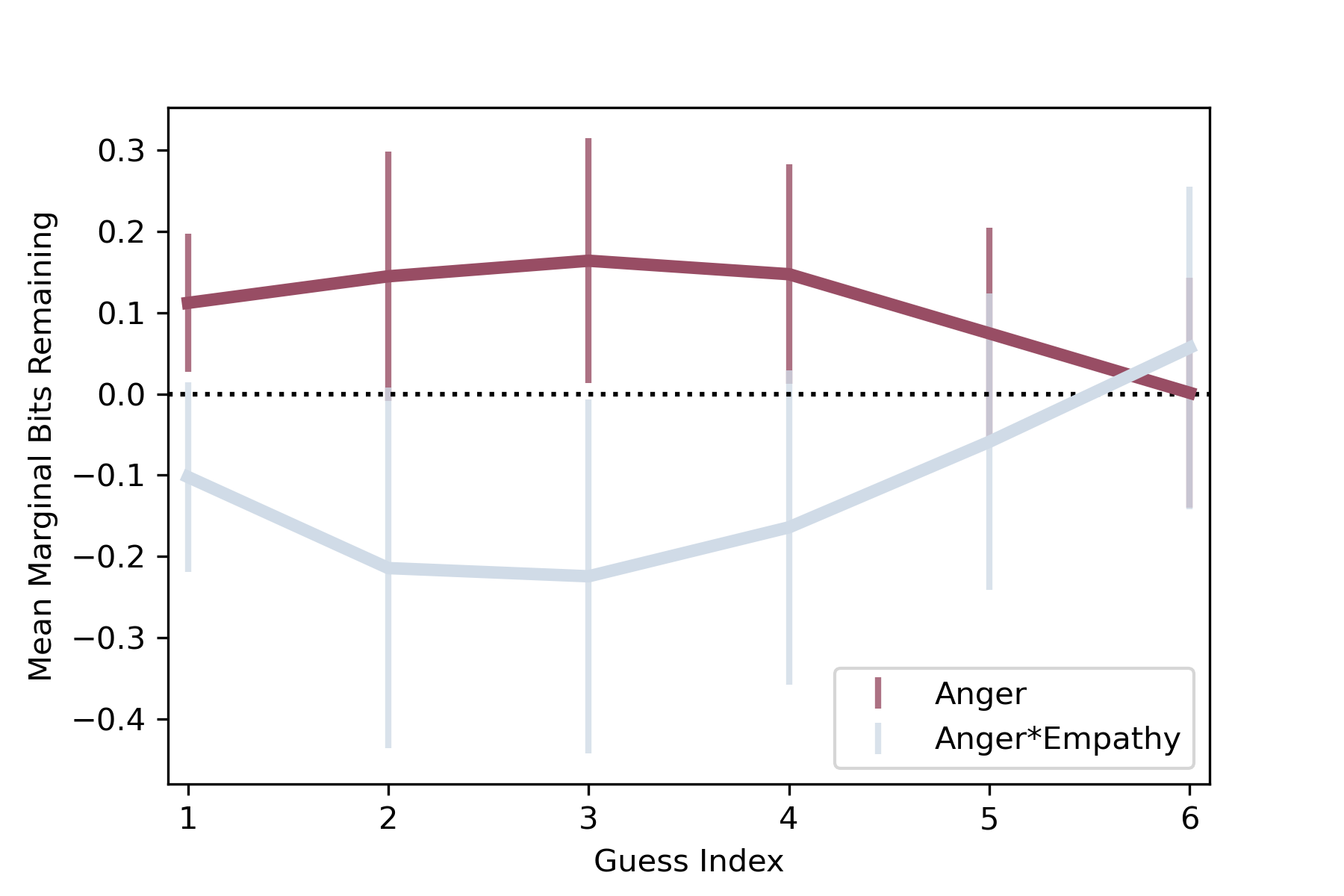}
\caption{Mean marginal bits remaining for all 2,315 possible Wordle solutions after each guess based on OLS regressions with robust standard errors clustered at the participant level. Bits are computed as $log_{2}(w)$ where w is the mean number of words remaining out of the 2,315 solutions. Error bars represent 95\% confidence intervals.}
\label{guess_index_2k}
\end{figure}

In Figure~\ref{guess_index_2k}, we present the 95\% confidence intervals for treatment effects on the mean marginal bits remaining from the 2,315 solutions for each guess iteration. We find that participants assigned to the anger elicitation condition have 0.11 to 0.16 additional bits of information remaining in their first four guesses ($p=0.01$, $p=0.06$, $p=0.03$, $p=0.03$) compared to the participants assigned to the control elicitation condition. In contrast, the Anger * Empathy interaction term ranges from -0.22 to -0.10 for the first four guesses ($p=0.09$, $p=0.06$, $p=0.04$, $p=0.10$). By the fifth and sixth guesses many participants have already identified the word and the average remaining bits is 0.6 and 0.47, respectively, so the lack of statistical differences across the anger elicitation and anger elicitation paired with computational empathy interventions in the fifth and sixth guesses can be explained by differential dropout and floor effects. 

As a robustness check, we also examine the treatment effects on the mean marginal bits remaining from the 12,972 valid words. We find that participants assigned to the anger elicitation condition have 0.13 to 0.17 additional bits of information remaining (based on the 12,972 valid words) in their first four guesses ($p=0.003$, $p=0.13$, $p=0.10$, $p=0.16$) compared to the participants assigned to the control elicitation condition. In contrast, the Anger * Empathy interaction term ranges from -0.25 to -0.10 for the first four guesses ($p=0.08$, $p=0.08$, $p=0.18$, $p=0.24$).

\subsection*{Self-Reported Affect and Additional Outcomes}

Based on OLS regressions of treatment effects following Equation 1 on self-reported valence and arousal that participants provided after completing the four rounds of Wordle, we find the anger elicitation has a statistically significant, negative effect on self-report affect. On a scale from 0 to 100, participants assigned to the anger condition report a 5.1 point lower arousal ($p=0.030$) and a 4.6 point lower valence ($p=0.064$) relative to participants assigned to the control emotion elicitation condition. We do not find statistically significant effects on self-reported valence and arousal from assignment to the computational empathy condition or the interaction between anger and computational empathy.

Based on OLS regressions of treatment effects on additional outcomes measured at the guess-level following Equation 1 including whether participants engaged in additional rounds of Wordle, guess sentiment based on the VADER rule-based model, the response time between guesses, the word frequency of participants' guesses, and the number of invalid attempts, we do not find statistically significant treatment effects of anger, empathy, or the interaction of anger and empathy. While we did not find significant treatment effects on these additional outcomes, we do see variation across participants across these features. 17\% of participants participated in at least one bonus round. Participants' guesses were classified by VADER as neutral for 86\% of words, positive for 7\% of words, and negative for 7\% of words. The mean time between each guess was 35 seconds. Participants submitted 6,074 5-letter strings, 3,175 unique valid words, and 1,609 unique words that are possible Wordle solutions, and 1,045 different first guesses of which 727 were valid Wordle solutions. Finally, 77\% of participants submitted at least 1 invalid guess.

\section*{Discussion}

How do incidental anger and computational empathy influence creative problem solving in a word guessing game? 

The results from our pre-registered experiment corroborate past research finding anger impairs decision-making and reduces depth of cognitive processing~\cite{lerner_beyond_2000,lerner_heart_2004, lerner_portrait_2006, lerner_emotion_2015}. In particular, we find the anger elicitation condition (relative to the control elicitation condition) leads participants to lose more often, make more guesses (and adjusted guesses), and submit less informative guesses (as measured by entropy reduction); these results are statistically significant. 

In contrast, we find the computational empathy intervention counteracts the negative effect of anger on performance. Participants assigned to both the anger elicitation condition and the empathic virtual agent perform better than participants assigned to the anger elicitation condition and the control virtual agent on all performance metrics, which are statistically significant at the $p<0.05$ level for round-level performance metrics and statistically significant at the $p<0.10$ level for guess-level entropy reduction. While computational empathy counteracts the negative effects of anger on cognitive performance, we do not find computational empathy changes performance for participants assigned to the control elicitation condition.

We find experience playing Wordle and depth of reflective reasoning as proxied by the CRT is strongly associated with performance, but we do not find significant heterogeneous treatment effects based on either participants' experience playing Wordle or participants' depth of reflective reasoning. The lack of heterogeneous treatment effects on both these characteristics and also participants' sex suggest that none of these characteristics make participants more or less vulnerable to the negative effects of anger or to the counterbalancing effects of computational empathy.

While anger and empathy influence overall performance, we do not find treatment effects on other outcomes like sentiment of guesses, response time between guesses, word frequency of guesses, number of invalid guesses, or whether participants engaged in additional rounds of Wordle after collecting payment for participating. The lack of treatment effects on these outcomes narrows possible mechanisms by which anger and empathy influence cognitive performance. 

After participants complete four rounds in the Affective Wordle Lab, we present participants an opportunity to self-report their valence and arousal with an affective slider~\cite{betella2016affective}. We find effects of the anger elicitation condition but not the empathic virtual agent personality on both arousal and valence. As expected, we find lower valence and arousal in participants assigned to the anger elicitation condition relative to the control condition. This may be surprising because anger is associated with positive arousal in the circumplex model of affect~\cite{russell1980circumplex}, but recent research that clusters semantic emotion categories and affective dimensions reveal different variations of anger that include both high and low arousal~\cite{cowen_self-report_2017}. The persistent effects of anger on self-reported affect -- that last through multiple rounds of the Affective Wordle Lab and are not counteracted by the computational empathy intervention -- reveal the effectiveness of the reflective writing exercise from Small and Lerner (2008)~\cite{small_emotional_2008} for eliciting incidental anger.

\section*{Limitations}

We evaluate treatment effects of an anger elicitation intervention and computational empathy intervention in a 2x2 factorial design on performance in an online word guessing game that involves creative problem solving. Similar to Duncker's Candle task and the RAT, this game represents only one kind of creative problem solving and it does not represent all creative problem solving. Moreover, we focused this experiment on incidental anger because it is more straightforward to elicit in experimental settings than integral anger. Future work may consider how integral emotions influence both creative problem solving and the ability of computational empathy to counteract the negative influence of anger.

In this experiment, we avoid precisely defining computational empathy and treat computational empathy as a gestalt of contextualized interactions involving affective mimicry and perspective taking. This gestalt treatment allows us to operationalize computational empathy as an intervention, but it prevents us from identifying the precise features that help counteract the negative effects of the anger elicitation intervention. Future research may consider the effectiveness of the empathic virtual agent without affective mimicry or without perspective taking or without some contexts to identify the most effective component parts and combination of component parts of computational empathy for improving an angry individual's performance. Likewise, future research may consider how computational empathy influences emotion regulation. 

\section*{Contributions and Implications}

We present a conceptual replication of experiments on anger and decision-making, and we corroborate previous findings that anger inhibits problem solving. Moreover, we present experimental evidence that computational empathy can counteract the negative effects of anger on creative problem solving. The countervailing force of computational empathy on anger highlights the importance of designing empathy into virtual agents to not only make people feel cared for but to boost people's creative performance.

Affective Wordle Lab presents a new tool and paradigm for interweaving a virtual agent within the constrained context of a game such that researchers can experimentally elicit emotions and manipulate virtual agents to evaluate computational empathy not only based on self-reports of how people feel but also as an assistive technology that can influence human decision-making and creative problem solving.

\section*{Data and Code Availability}

We open-sourced the code for Affective Wordle Lab at \href{https://github.com/MITMediaLabAffectiveComputing/WordleLab}{https://github.com/MITMediaLabAffectiveComputing/WordleLab} and share anonymized participant data and replication code at \href{https://github.com/mattgroh/affective_wordle_lab_replication}{https://github.com/mattgroh/affective\_wordle\_lab\_replication}.

\section*{Acknowledgments}

We acknowledge JcToon on Rive for creating and licensing the Animated Login Screen. 
We thank Neska El Haouij and Ila Kumar for feedback on an early version of the experiment and Boyu Zhang for feedback on an early draft.

\section*{Ethics and Informed Consent}

This research complies with all relevant ethical regulations. 
The Massachusetts Institute of Technology’s Committee on the Use of Humans as Experimental Subjects determined this study to fall under Exempt Category 3: Benign Behavioral Intervention and Exempt Category 2: Educational Testing, Surveys, Interviews or Observation with id E-3888. 

All participants are informed that ``WordleLab is a research project created by 
the MIT Media Lab'' and ``All submissions are collected anonymously for research purposes, and participation is entirely voluntary. For questions, please contact 
wordlelab@media.mit.edu.'' All participants are recruited from the Prolific survey platform with the following message: ``The Wordle Lab Experiment is a research project created at the  MIT Media Lab to study how people play Wordle. First, we'll ask you to share three to five examples of something relevant to you, then you'll play 4 rounds of Wordle, and last we'll ask you a few follow up questions. All data is collected anonymously. We estimate this experiment to take about 15 minutes.'' We compensated participants with \$2.38 each, which is a rate of \$9.52 an hour.

\printbibliography
\end{document}


\section*{Supplementary Information: Empathic Virtual Agents Facilitate Creative Problem Solving}

\begin{table}[!htbp] \centering
\begin{tabular}{@{\extracolsep{5pt}}lccc}
\\[-1.8ex]\hline
\hline \\[-1.8ex]
\\[-1.8ex] & \multicolumn{1}{c}{Did Win} & \multicolumn{1}{c}{Guesses} & \multicolumn{1}{c}{Guesses (Adjusted)}  \\
\hline \\[-1.8ex]
 Constant & 0.55$^{***}$ & 5.17$^{***}$ & 5.62$^{***}$ \\
  & (0.04) & (0.08) & (0.11) \\
 Anger & -0.08$^{}$ & 0.15$^{}$ & 0.23$^{}$ \\
  & (0.06) & (0.11) & (0.16) \\
 Empathy & 0.04$^{}$ & 0.01$^{}$ & -0.03$^{}$ \\
  & (0.06) & (0.11) & (0.16) \\
 Anger * Empathy & 0.05$^{}$ & -0.16$^{}$ & -0.21$^{}$ \\
  & (0.08) & (0.16) & (0.23) \\
 CRT & 0.09$^{***}$ & -0.20$^{***}$ & -0.29$^{***}$ \\
  & (0.02) & (0.04) & (0.05) \\
 CRT * Anger & 0.02$^{}$ & -0.02$^{}$ & -0.03$^{}$ \\
  & (0.02) & (0.05) & (0.07) \\
 CRT * Empathy & -0.03$^{}$ & 0.05$^{}$ & 0.08$^{}$ \\
  & (0.02) & (0.05) & (0.07) \\
 CRT * Anger * Empathy & 0.01$^{}$ & -0.01$^{}$ & -0.02$^{}$ \\
  & (0.03) & (0.07) & (0.10) \\
\hline \\[-1.8ex]
 Observations & 3,975 & 3,975 & 3,975 \\
  Number of Participants & 1006 & 1006 & 1006 \\

\hline
\hline \\[-1.8ex]
\textit{Note:} & \multicolumn{3}{r}{$^{*}$p$<$0.05; $^{**}$p$<$0.01; $^{***}$p$<$0.001} \\
\end{tabular}
  \caption{Heterogeneity of Treatment Effects: OLS regressions examining heterogeneous treatment effects on performance of the Cognitive Reflection Test. Robust standard errors are clustered on at the participant level.}
\end{table}

\begin{table}[!htbp] \centering
\begin{tabular}{@{\extracolsep{5pt}}lccc}
\\[-1.8ex]\hline
\hline \\[-1.8ex]
\\[-1.8ex] & \multicolumn{1}{c}{Did Win} & \multicolumn{1}{c}{Guesses} & \multicolumn{1}{c}{Guesses (Adjusted)}  \\
\hline \\[-1.8ex]
 Constant & 0.78$^{***}$ & 4.71$^{***}$ & 4.93$^{***}$ \\
  & (0.02) & (0.05) & (0.07) \\
 Anger & -0.06$^{}$ & 0.11$^{}$ & 0.17$^{}$ \\
  & (0.03) & (0.07) & (0.09) \\
 Empathy & -0.00$^{}$ & 0.07$^{}$ & 0.07$^{}$ \\
  & (0.03) & (0.07) & (0.10) \\
 Anger * Empathy & 0.07$^{}$ & -0.20$^{}$ & -0.27$^{}$ \\
  & (0.04) & (0.10) & (0.14) \\
 Never Played Wordle & -0.20$^{***}$ & 0.35$^{***}$ & 0.55$^{***}$ \\
  & (0.04) & (0.10) & (0.14) \\
 Never Played Wordle * Anger & 0.00$^{}$ & 0.07$^{}$ & 0.06$^{}$ \\
  & (0.06) & (0.13) & (0.18) \\
 Never Played Wordle * Empathy & -0.05$^{}$ & 0.11$^{}$ & 0.15$^{}$ \\
  & (0.06) & (0.14) & (0.19) \\
  Never Played Wordle * Anger * Empathy & -0.01$^{}$ & 0.04$^{}$ & 0.06$^{}$ \\
  & (0.09) & (0.19) & (0.26) \\
\hline \\[-1.8ex]
 Observations & 3,975 & 3,975 & 3,975 \\
 Number of Participants & 1006 & 1006 & 1006 \\

\hline
\hline \\[-1.8ex]
\textit{Note:} & \multicolumn{3}{r}{$^{*}$p$<$0.05; $^{**}$p$<$0.01; $^{***}$p$<$0.001} \\
\end{tabular}
  \caption{Heterogeneity of Treatment Effects: OLS regressions examining heterogeneous treatment effects on performance of participants' experience playing Wordle. Robust standard errors are clustered on at the participant level.}
\end{table}

\begin{table}[!htbp] \centering
\begin{tabular}{@{\extracolsep{5pt}}lccc}
\\[-1.8ex]\hline
\hline \\[-1.8ex]
\\[-1.8ex] & \multicolumn{1}{c}{Did Win} & \multicolumn{1}{c}{Guesses} & \multicolumn{1}{c}{Guesses (Adjusted)}  \\
\hline \\[-1.8ex]
 Constant & 0.70$^{***}$ & 4.82$^{***}$ & 5.12$^{***}$ \\
  & (0.03) & (0.07) & (0.10) \\
 Anger & -0.06$^{}$ & 0.18$^{*}$ & 0.24$^{*}$ \\
  & (0.04) & (0.09) & (0.13) \\
 Empathy & -0.04$^{}$ & 0.20$^{**}$ & 0.24$^{*}$ \\
  & (0.04) & (0.10) & (0.13) \\
 Anger * Empathy & 0.09$^{}$ & -0.26$^{**}$ & -0.35$^{*}$ \\
  & (0.06) & (0.13) & (0.18) \\
 Female & 0.03$^{}$ & -0.01$^{}$ & -0.04$^{}$ \\
  & (0.04) & (0.09) & (0.12) \\
 Female * Anger & -0.01$^{}$ & -0.07$^{}$ & -0.06$^{}$ \\
  & (0.06) & (0.13) & (0.17) \\
 Female * Empathy & 0.04$^{}$ & -0.18$^{}$ & -0.22$^{}$ \\
  & (0.06) & (0.13) & (0.18) \\
 Female * Anger * Empathy & -0.02$^{}$ & 0.12$^{}$ & 0.14$^{}$ \\
  & (0.08) & (0.18) & (0.25) \\

\hline \\[-1.8ex]
 Observations & 3,975 & 3,975 & 3,975 \\
 Number of Participants & 1006 & 1006 & 1006 \\

\hline
\hline \\[-1.8ex]
\textit{Note:} & \multicolumn{3}{r}{$^{*}$p$<$0.05; $^{**}$p$<$0.01; $^{***}$p$<$0.001} \\
\end{tabular}
  \caption{Heterogeneity of Treatment Effects: OLS regressions examining heterogeneous treatment effects on performance of participants' reported sex. Robust standard errors are clustered on at the participant level.}
\end{table}

\begin{table}[!htbp] \centering
\begin{tabular}{@{\extracolsep{5pt}}lcccccc}
\\[-1.8ex]\hline
\hline \\[-1.8ex]
& \multicolumn{6}{c}{\textit{Dependent variable: Mean Bits Remaining from 2,315 possible solutions}} \
\cr \cline{6-7}
\\[-1.8ex] & \multicolumn{1}{c}{1st Guess} & \multicolumn{1}{c}{2nd Guess} & \multicolumn{1}{c}{3rd Guess} & \multicolumn{1}{c}{4th Guess} & \multicolumn{1}{c}{5th Guess} & \multicolumn{1}{c}{6th Guess}  \\
\hline \\[-1.8ex]
 Constant & 7.0351$^{***}$ & 3.5355$^{***}$ & 1.6630$^{***}$ & 0.9101$^{***}$ & 0.5916$^{***}$ & 0.4711$^{***}$ \\
  & (0.0305) & (0.0537) & (0.0520) & (0.0475) & (0.0479) & (0.0560) \\
 Anger & 0.1120$^{**}$ & 0.1447$^{}$ & 0.1638$^{*}$ & 0.1473$^{*}$ & 0.0745$^{}$ & 0.0014$^{}$ \\
  & (0.0433) & (0.0781) & (0.0769) & (0.0689) & (0.0664) & (0.0720) \\
 Empathy & 0.0102$^{}$ & 0.0541$^{}$ & 0.1343$^{}$ & 0.0527$^{}$ & -0.0319$^{}$ & -0.0290$^{}$ \\
  & (0.0420) & (0.0777) & (0.0773) & (0.0686) & (0.0663) & (0.0737) \\
 Anger * Empathy & -0.1023$^{}$ & -0.2143$^{}$ & -0.2244$^{*}$ & -0.1642$^{}$ & -0.0590$^{}$ & 0.0569$^{}$ \\
  & (0.0595) & (0.1132) & (0.1111) & (0.0986) & (0.0930) & (0.1012) \\

\hline \\[-1.8ex]
 Observations & 3,981 & 3,968 & 3,850 & 3,339 & 2,553 & 1,760 \\
  Number of Participants & 1006 & 1006 & 1006 & 1005 & 982 & 838 \\
\hline
\hline \\[-1.8ex]
\textit{Note:} & \multicolumn{6}{r}{$^{*}$p$<$0.05; $^{**}$p$<$0.01; $^{***}$p$<$0.001} \\
\end{tabular}
\caption{Guess Level Entropy Reduction: OLS regressions examining entropy reduction at the guess level based on the 2,315 possible Wordle solutions. Robust standard errors are clustered on at the participant level.}
\end{table}

\begin{table}[!htbp] \centering

\begin{tabular}{@{\extracolsep{5pt}}lcccccc}
\\[-1.8ex]\hline
\hline \\[-1.8ex]
& \multicolumn{6}{c}{\textit{Dependent variable: Mean Bits Remaining from 12,972 possible words}} \
\cr \cline{6-7}
\\[-1.8ex] & \multicolumn{1}{c}{1st Guess} & \multicolumn{1}{c}{2nd Guess} & \multicolumn{1}{c}{3rd Guess} & \multicolumn{1}{c}{4th Guess} & \multicolumn{1}{c}{5th Guess} & \multicolumn{1}{c}{6th Guess}  \\
\hline \\[-1.8ex]
 Constant & 9.0903$^{***}$ & 5.3546$^{***}$ & 3.0549$^{***}$ & 2.1170$^{***}$ & 1.6589$^{***}$ & 1.3313$^{***}$ \\
  & (0.0312) & (0.0682) & (0.0720) & (0.0665) & (0.0715) & (0.0862) \\
 Anger & 0.1255$^{**}$ & 0.1468$^{}$ & 0.1745$^{}$ & 0.1362$^{}$ & 0.0549$^{}$ & -0.0118$^{}$ \\
  & (0.0428) & (0.0985) & (0.1073) & (0.0960) & (0.0984) & (0.1141) \\
 Empathy & 0.0163$^{}$ & 0.0875$^{}$ & 0.1330$^{}$ & 0.0368$^{}$ & -0.0665$^{}$ & -0.0220$^{}$ \\
  & (0.0420) & (0.0975) & (0.1058) & (0.0954) & (0.0978) & (0.1156) \\
 Anger * Empathy & -0.1021$^{}$ & -0.2510$^{}$ & -0.2035$^{}$ & -0.1608$^{}$ & -0.1264$^{}$ & -0.0294$^{}$ \\
  & (0.0587) & (0.1419) & (0.1534) & (0.1371) & (0.1379) & (0.1600) \\

\hline \\[-1.8ex]
 Observations & 3,981 & 3,968 & 3,850 & 3,339 & 2,553 & 1,760 \\
 Number of Participants & 1006 & 1006 & 1006 & 1005 & 982 & 838 \\
\hline
\hline \\[-1.8ex]
\textit{Note:} & \multicolumn{6}{r}{$^{*}$p$<$0.05; $^{**}$p$<$0.01; $^{***}$p$<$0.001} \\
\end{tabular}
  \caption{Guess Level Entropy Reduction: OLS regressions examining entropy reduction at the guess level based on the 12,972 possible Wordle words. Robust standard errors are clustered on at the participant level.}
\end{table}

\begin{table}[!htbp] \centering
\begin{tabular}{@{\extracolsep{5pt}}lccc}
\\[-1.8ex]\hline
\hline \\[-1.8ex]
\\[-1.8ex] & \multicolumn{1}{c}{Arousal} & \multicolumn{1}{c}{Valence} & \multicolumn{1}{c}{Started Bonus Rounds}  \\
\hline \\[-1.8ex]
 Constant & 61.05$^{***}$ & 64.54$^{***}$ & 0.22$^{***}$ \\
  & (1.58) & (1.75) & (0.03) \\
 Anger & -5.10$^{*}$ & -4.60$^{}$ & -0.01$^{}$ \\
  & (2.35) & (2.48) & (0.04) \\
 Empathy & -1.13$^{}$ & 1.46$^{}$ & 0.03$^{}$ \\
  & (2.26) & (2.36) & (0.04) \\
 Anger * Empathy & 4.48$^{}$ & 1.64$^{}$ & -0.05$^{}$ \\
  & (3.27) & (3.34) & (0.05) \\
\hline \\[-1.8ex]
 Observations & 1,006 & 1,006 & 1,006 \\
  Number of Participants & 1006 & 1006 & 1006 \\

\hline
\hline \\[-1.8ex]
\textit{Note:} & \multicolumn{3}{r}{$^{*}$p$<$0.05; $^{**}$p$<$0.01; $^{***}$p$<$0.001} \\
\end{tabular}
  \caption{Self-Report Affect: OLS regressions examining self-reported arousal and valence and playing additional rounds after collecting payment. Robust standard errors are clustered on at the participant level.}

\end{table}

\begin{table}[!htbp] \centering
\scalebox{0.9}{
\begin{tabular}{@{\extracolsep{5pt}}lcccccc}
\\[-1.8ex]\hline
\hline \\[-1.8ex]
\\[-1.8ex] & \multicolumn{1}{c}{Frequency} & \multicolumn{1}{c}{Response Time} & \multicolumn{1}{c}{Positive} & \multicolumn{1}{c}{Neutral} & \multicolumn{1}{c}{Negative} & \multicolumn{1}{c}{Valid}  \\
\hline \\[-1.8ex]
 Contstant & 0.5509$^{***}$ & 0.5737$^{***}$ & 0.0725$^{***}$ & 0.8559$^{***}$ & 0.0717$^{***}$ & 0.8365$^{***}$ \\
  & (0.0330) & (0.0248) & (0.0043) & (0.0055) & (0.0043) & (0.0134) \\
 Anger & -0.0108$^{}$ & -0.0316$^{}$ & -0.0054$^{}$ & 0.0056$^{}$ & -0.0002$^{}$ & -0.0078$^{}$ \\
  & (0.0439) & (0.0325) & (0.0060) & (0.0079) & (0.0063) & (0.0186) \\
 Empathy & 0.0043$^{}$ & -0.0109$^{}$ & -0.0047$^{}$ & 0.0086$^{}$ & -0.0039$^{}$ & -0.0019$^{}$ \\
  & (0.0439) & (0.0344) & (0.0063) & (0.0083) & (0.0060) & (0.0175) \\
 Anger * Empathy & -0.0091$^{}$ & 0.0188$^{}$ & 0.0054$^{}$ & -0.0067$^{}$ & 0.0013$^{}$ & -0.0406$^{}$ \\
  & (0.0602) & (0.0481) & (0.0086) & (0.0117) & (0.0089) & (0.0292) \\
\hline \\[-1.8ex]
 Observations & 19,451 & 22,675 & 19,451 & 19,451 & 19,451 & 23,681 \\
  Number of Participants & 1006 & 1006 & 1006 & 1006 & 1006 & 1006 \\

\hline
\hline \\[-1.8ex]
\textit{Note:} & \multicolumn{6}{r}{$^{*}$p$<$0.05; $^{**}$p$<$0.01; $^{***}$p$<$0.001} \\
\end{tabular}}
  \caption{Additional Outcomes: OLS regressions examining additional outcomes including word frequency, response time (in minutes), sentiment (positive, neutral, negative), and frequency of valid guesses. Robust standard errors are clustered on at the participant level.}
\end{table}